\def \bfr {\begin{flushright}}
\def \efr {\end{flushright}}
\def \caja {\makebox[3.2cm][1]}
\def \L {{\cal L}}
\def \R {{\it R}}
\def \d {\hbox{d}\,}
\def \sgn {\hbox{sgn}}
\def \e {\hbox{e}}
\def \l {{\Lambda\over8}}
\def \v {\vskip}
\def \pp {\partial_+}
\def \mm {\partial_-}
\def \Re {\hbox{Re}}

\def \be {\begin{equation}}
\def \ee {\end{equation}}

\def \ba {\begin{array}}
\def \ea {\end{array}}

\def \bea {\begin{eqnarray}}
\def \eea {\end{eqnarray}}

\def \bfr {\begin{flushright}}
\def \efr {\end{flushright}}
\def \caja {\makebox[3.2cm][1]}
\def \square {\hbox{$\sqcup\!\!\!\!\sqcap$}}

\def \pifi {\pi_\Phi}
\def \pia {\pi_a}
\def \C {{\cal C}}
\def \N {{\cal N}}
\def \C {{\cal C}}
\def \J {{\cal J}}

\documentstyle[a4,12pt]{article}

\begin{document}

%%%%%%%%%%%%%%%%%%%%%%%%%%%%%%%%%%%%%%%%%%%%%%%%%%%%%%%%%%%%%%
\pagestyle{empty}
\bfr
\caja{{\bf FTUV/93-15}}\\
\caja{{\bf IFIC/93-10}}\\
\caja{{\bf Imperial-TP/93-94/18}}
\efr
\v 1cm

\vfill

\begin{center}
{\bf ON THE REDUCED CANONICAL QUANTIZATION
OF THE INDUCED
2D-GRAVITY}\footnote[2]{
Work partially supported by the C.I.C.Y.T. and the D.G.I.C.Y.T.}
\v 0.3cm
Jos\'e Navarro-Salas$^{1,2}$, Miguel Navarro$^{2,3,4}$,\\
C.F. Talavera$^{1,2}$ and V\'\i ctor Aldaya$^{2,4}$
\v 0.3cm
\end{center}

\vfill

\noindent 1.- Departamento  de  F\'\i sica  Te\'orica, Burjassot-46100,
Valencia, Spain.

\noindent 2.- IFIC, Centro Mixto Universidad de
Valencia-CSIC, Burjassot-46100, Valencia, Spain.

\noindent 3.- The Blackett Laboratory, Imperial College, London SW7 2BZ,
United Kingdom.

\noindent 4.- Instituto Carlos I de F\'\i sica Te\'orica y Computacional,
Facultad  de  Ciencias, Universidad de Granada, Campus de Fuentenueva,
18002, Granada, Spain.

\vfill

\begin{center}{\bf Abstract}
\end{center}

The quantization of the induced 2d-gravity on a compact spatial section
is carried out in three different ways.
In the three approaches the supermomentum constraint is solved
at the classical level but they differ in the way the hamiltonian
constraint is imposed.
We compare these approaches establishing an isomorphism between the
resulting Hilbert spaces.

\vfill
\eject

%%%%%%%%%%%%%%%%%%%%%%%%%%%%%%%%%%%%%%%%%%%%%%%%%%%%%%%%%%%%%%

\setcounter{page}{1}
\pagestyle{plain}

\section{Introduction}
\label{a}

Generally covariant theories in a two-dimensional space-time collect the
advantages of both being much simpler than the corresponding
theories in 3+1 and 2+1 dimensions, and of having a sufficiently
rich structure which can shed light on the issues that appear in
quantizing higher dimensional theories.

Several years ago Jackiw and Teitelboim \cite{[1],[2]} proposed
the equation

\be R + \frac\Lambda2 = 0\ee
as the natural analogue of the vacuum Einstein equations with a
cosmological term. This equation can be obtained form a local
variational principle if a scalar field, playing the role of a
lagrangian multiplier, is incorporated in the theory. The above
equation can also be derived from the induced 2d gravity
\cite{[3]}:
\begin{eqnarray}
S = \frac{c}{96\pi}\int\,\sqrt{-g} (R{\square}^{-1}R + \Lambda) \; .
\label{1}
\end{eqnarray}
This action is non-local but it is preferable to convert it into
a local one by introducing an auxiliary scalar field $\Phi$. The
action can be written as

\begin{eqnarray}S=\frac{1}{2}\int\sqrt{-g}(g^{\mu\nu}\partial_\mu\Phi\partial_\nu\Phi +
2R\Phi + \Lambda)\;.\label{action1}\end{eqnarray}

The aim of this paper is to carry out a canonical analysis of the
induced 2d-gravity theory (\ref{1}) in three different
ways (we shall restrict ourselves to the case of a compact
spatial section). The first one is presented in section 3 and it
was  spelled out in \cite{[4]}. It is based on the
covariant formulation of the canonical formalism \cite{[5]}.
The reduced phase space of the theory turns
out to be a two-dimensional cotangent bundle, and the corresponding
(geometric) quantization permits to determine the Hilbert space.
In section 3 we introduce the ADM formulation of the theory. By
gauge fixing and imposing the supermomentum
constraint we can reduce the theory to a finite-dimensional
system. At this point one can choose different ways to quantize
the theory. One way is to look for the reduced hamiltonian (this
requires a complete gauge fixing) and then to impose the
corresponding Sch\"odinger equation. This is our second
approach and it is developed in section 4. The third approach
(section 5) is based on the (reduced) Wheeler-DeWitt equation.
Along the paper we set up the equivalence of these approaches
establishing an isomorphism between the corresponding Hilbert
spaces.

\section{Covariant phase-space quantization}
\label{b}

The covariant definition of the reduced phase-space
\cite{[5]}
has been very useful to determine the phase space of
a variety of field theories \cite{[6],[7],[8]}. In this
approach, the reduced phase space is defined as the set
of all solutions of the classical theory, modulo gauge
transformations (see below). The symplectic form is defined as
follows.

Let us consider a field theory with fields $\Psi^\alpha$ and Lagrangian
$\L$. If we vary the fields in the Lagrangian we get
%%2
\begin{eqnarray}\delta\L = \partial_\mu \jmath^\mu + (E-
L)_\alpha\delta\Psi^\alpha\;.\label{2}\end{eqnarray}

If we regard now $\delta$ as an exterior derivative operator in
the space of classical solutions $\Psi^\alpha(x)$, we can, in a natural
way,  pullback (\ref{2}) to the space of all solutions of the
equations of motion, (E-L)$=0$, and consider $\jmath^\mu$ as a
vector-valued one-form on this space.

Since $\delta\jmath^\mu$ is a conserved current,
$\partial_\mu\delta\jmath^\mu = 0$, it is natural to define the
(pre-)symplectic form $\omega$ as the corresponding conserved
charge,

\be\omega = -\int_\Sigma\,\delta\jmath^\mu\d
\sigma_\mu\label{3}\ee
($\Sigma$ is any spatial hypersurface of the space-time).
With this definition, we prevent $\omega$ from depending on
$\Sigma$ or on the time co-ordinate. Because $\delta\jmath^\mu$ is
exact, so is $\omega$ and thus closed.

The only property we can not assure for $\omega$ is nondegenerateness
since $\omega$ as defined above can have a non-trivial kernel. We
define now the gauge transformations as the ones generated by the
kernel of $\omega$. If now, in the space of all
solutions, we take modulus by the gauge transformations we get a
symplectic space which is called the reduced (or physical)
phase space.

Let us apply the program above to the induced 2d-gravity.
The equations of motion obtained from (\ref{1}) imply the
vanishing of the stress-tensor $T_{\mu\nu}$ that is given by

\begin{equation} \begin{array}{lcl}
T_{\mu\nu}& =  -&\nabla_\mu\Phi\nabla_\nu\Phi+
2\nabla_\mu\nabla_\nu\Phi + \frac{1}{ 2}g_{\mu\nu}\nabla^\alpha
\Phi\nabla_\alpha\Phi   \\
& &-g_{\mu\nu}(2R +
\frac{1}{2}\Lambda)\>,\end{array}\label{tensor1}\end{equation}
and the relation of $\Phi$ with the curvature:
\be \square \Phi = R\>.\label{motionPhi1}\ee

If we use now the gauge invariance of the metric under diffeomorphisms to
bring it to a conformally flat form,

\be\d\hbox{s}^2 = -2\e^\rho\d x^+\d x^-\>,\label{flatmetric}\ee
($x^+= t+x,\>x^-=t-x$ are the light-cone coordinates)
the equations of motion split into the relation of $\Phi$ with
the curvature,
\be
\square\Phi\equiv2e^{-\rho}\pp\mm\Phi
=R\equiv-2\e^{-\rho}\pp\mm \rho\>,\label{motionPhi2}\ee
the Liouville equation,
\be0=T_{+-} = 2\pp\mm \rho -
\frac{\Lambda}{2}\e^{\rho}\>,
\label{Liouville}\ee
and the constraints
\begin{eqnarray}  0=T_{++} &=& -(\partial_+\Phi)^2 + 2\partial_+^2\Phi -
2\partial_+\rho\partial_+\Phi = 0 \label{(4.21a)}\;,   \\
0=T_{--} &=& -(\partial_-\Phi)^2 + 2\partial_-^2\Phi -
2\partial_-\rho\partial_-\Phi = 0\;. \label{constraints}\end{eqnarray}

The general solution for the metric field $g_{\mu\nu}$ and the
dilaton field $\Phi$ can be easily found \cite{[4]}, and
can be written as follows:
%%4
\begin{eqnarray}
\d\hbox{s}^2 = -2\frac{\partial_+A\,\partial_-B}{ \left(1 - \l
AB\right)^2}\d x^+\d x^-,\label{metric}\end{eqnarray}

\begin{eqnarray}\Phi = \ln \lambda\frac{(1- \l AB)^2}{ \left((d-a)A +
b\right)^2{(\l B)}^2}\;,\label{Phi1}\end{eqnarray}
and

\begin{eqnarray}\Phi = \ln \lambda \frac{(1- \l AB)^2}{ \left(-b\l B +(d-
a)\right)^2}\>,\label{Phi2}\end{eqnarray}
where  $a, b, d$ are such that
$M=\left(\begin{array}{cc}a&b   \\ 0&d\end{array}
\right)$ belong to the affine subgroup of $PSL(2,R)$,
i.e., $a=d^{-1}$, and $A=A(x^+), B=B(x^-)$
verify the monodromy transformations properties (we choose the length
of the circle equal to unity)
\begin{eqnarray}
 A(y + 1) & = &\frac{a A(y)
+ b}{d}\> \equiv M\left(
A(y)\right)\; ,\label{(3.15a)}   \\
-\l B(y - 1) & = &\frac{-d\l B(y)
}{ b\l B(y) + a} \equiv M^{-1T}\left(-\l B(y)
\right)\; .\label{7}
\end{eqnarray}

If we choose the spatial hypersurface as the one defined by $t=t_0$
the symplectic form can be written as:

\be   \omega = \int_x^{x+1}
\left[-\delta\Phi\delta(\partial_++\partial_-)(\Phi +\rho) +
\delta(\partial_+ +\partial_-)\Phi\delta\rho\right]\>.
\label{omega1}\ee

The projection onto the space of classical solutions takes
an special form

\be \omega = \frac{1}{2}\int_x^{x+1}\left(\partial_+
- \partial_-\right)W\>, \label{9}\ee
where $W$ is given by

\begin{eqnarray}
W& =  &\frac{1}{2}\{
\delta\ln \frac{\partial_+A}{\partial_-B}\left(\frac{\frac{\Lambda}{8} B
 }{(d-a) A +b}\right)^2\,
\delta\ln \lambda\frac{(1- \l AB)^2}{ ((d-a) A +b)^2
(\frac{\Lambda}{8} B)^2}\nonumber \\
& &+\delta\ln(1- \l AB)^2\,\delta\ln\frac{A}{ B}\left(\frac{\frac{\Lambda}{8} B
 }{(d-a) A +b}\right)^2 \label{W1}\\& &+
\delta\ln\frac{((d-a) A + b)^2}{(\frac{\Lambda}{8})^2}\,
\delta\ln \frac{(d-a)^2}{(\frac{\Lambda}{8}
B )^2}\}\>\; , \nonumber\end{eqnarray}
for the solution (\ref{Phi1}) for $\Phi$, and by
\begin{eqnarray}
W& = &\frac{1}{2}\{
\delta\ln \frac{\partial_+A}{\partial_-B}\left( b\frac{\Lambda}{8} B
-(d-a)\right)^2\,
\delta\ln \frac{(1- \l AB)^2}{
( b\frac{\Lambda}{8} B - (d-a))^2}\nonumber\\
& &+\delta\ln(1- \l AB)^2\,\delta\ln\frac{A}{ B}\left( b\frac{\Lambda}{8} B
-(d-a)\right)^2\label{W2}\\& &+
\delta\ln (b\frac{\Lambda}{8})^2\,
\delta\ln ( b\frac{\Lambda}{8}
B-(d-a))^2\}\>\; ,\nonumber\end{eqnarray}
for the solution (\ref{Phi2}) for $\Phi$.

In any case, since $\omega$ does not depend on the coordinate $x$
in (\ref{9}), it cannot depend on either of the functions $A$ or
$B$. So $\omega$ will depend only on the classes of monodromy
transformations to which the functions $A$ and $B$ belong, and on the
parameter $\lambda$ in (\ref{Phi1},\ref{Phi2}). Moreover, if we
transform the functions $A$ and $B$ as

\begin{eqnarray}  &A\longrightarrow h(A) \; ,\label{12}   \\
&-\l B\longrightarrow h^{-1T}(-\l B)\>\; ,\label{13}\end{eqnarray}
where $h$ is a constant affine matrix acting as a M\"obius transformation,
we get the same solution of the equations of motion. Under the
transformation (\ref{12},\ref{13}) the monodromy parameters
transform as

\begin{eqnarray}
M\longrightarrow hMh^{-1}\; .
\label{14}
\end{eqnarray}

Thus, two solutions which differs on a transformation of the type
(\ref{14}) are the same point of the reduced phase space. The
only invariant quantity under the transformation (\ref{14}), and
hence the only allowed monodromy dependence in $\omega$, is
the parameter $a$.

A direct computation from (\ref{W1}, \ref{W2}) leads to:
\begin{eqnarray}\omega=2\frac{\delta\lambda}{\lambda}\,\frac{\delta a}{
a}\>,\label{15}\end{eqnarray}
for the solutions in (\ref{W1}), and

\begin{eqnarray}\omega=-2\frac{\delta\lambda}{\lambda}\,\frac{\delta a}{
a}\>.\label{16}\end{eqnarray}
for the solutions in (\ref{W2}).

{}From (\ref{15},\ref{16}) and previous considerations, we are tempted
to assume that the reduced phase space is of the form

\be T^*\left(G/\hbox{ad} G\right)\cup T^*\left(G/\hbox{ad} G\right)
\>,\label{17}\ee
where $G$ is the affine subgroup of $PSL(2,\R)$.
This would lead us to a Hilbert
space of the form

\be{\cal H} ={\cal H}^{(+)}\oplus {\cal H}^{(-)},\label{18}\ee
where

\be {\cal H}^{(+)}= {\cal H}^{(-)} = L^2(\R^+)\oplus{\it
C}^2\>.\label{19}\ee

The result (\ref{18}), in which the Hilbert space has a
continuum and a discrete sector is in accordance with some results
obtained by BRST methods \cite{[9]}. However, there are
also some evidence that the discrete sector cannot be endowed with
a well defined inner product. This result is achieved here by
showing that, in fact, the discrete sector actually does not appear.
This is a consequence of the additional symmetry $A\rightarrow -
A,\>\>B\rightarrow-B,\>\>a\rightarrow a,\>\>b\rightarrow-b$ that
identify the otherwise distinct parabolic ($a=1$) solutions.

Let us write the classical solutions (\ref{metric},\ref{Phi1},\ref{Phi2})
in a more explicit form. To this end we should completely fix the
space-time coordinates by imposing and additional ``gauge-type"
condition. From (\ref{motionPhi2}) we observe that
\be
\pp\mm (\Phi+\rho) = 0\; .
\ee
Therefore we can (and we will) choose a spatially homogeneous conformal gauge
by imposing

\be \Phi + \rho = \epsilon + 2pt\>,\ee
where $\epsilon$ and $p$ are constant parameters.

If $a\neq1$, i.e., if the
monodromy class is hyperbolic, the solutions
take the form:

\begin{eqnarray}
\d \hbox{s}^2 = - 2\frac{8}{|\Lambda|}
\frac{p^2\e^{2{{pt}}}}{\left(1-\sgn\Lambda\e^{2{{pt}}}\right)^2}\d
x^+\d x^-\label{101}\; ,\\
\Phi=\ln\lambda\frac{\left(1-\sgn\Lambda\e^{{{2pt}}}\right)^2}
{4(\sinh p/2)^2\e^{4{{pt}}}}\; ,\label{102}\\
\Phi=\ln\lambda\frac{\left(1-\sgn\Lambda\e^{2{{pt}}}\right)^2}
{4(\sinh\frac{p}{2})^2}\; ,\label{103}\end{eqnarray}
where $\e^p = a = d^{-1}$.  On the other hand if $a=1=d$, the
unique parabolic solution takes the form

\begin{eqnarray}
\d \hbox{s}^2& = -&\frac{2}{\Lambda}\frac{1}{t^2}\d x^+\d x^-
\label{104}\; ,\\
\Phi& = & \ln 4\lambda {t^2}\; .\label{105}\end{eqnarray}

We can easily see that (\ref{102}) and (\ref{103}) transform
into each other when we make the replacement $a\leftrightarrow a^{-1}=d$,
under which also (\ref{15}) and (\ref{16}) transform into each
other.
Moreover, when $\Lambda>0$, (\ref{104}) has the right signature
and can be obtained from (\ref{101}-\ref{103}) in the limit
$p\rightarrow0$ ($a\rightarrow1$). So we can conclude
that the phase space for $\Lambda>0$ is just
\be T^*(\R)\cup T^*(\R) \; ,\label{108}\ee
with the symplectic form
\begin{equation} \omega = 2\delta(\ln \lambda)\>\delta
p.\label{109}\end{equation}
The two sectors in (\ref{108}) correspond to whether the scalar
field is expanding or contracting.

The cotangent bundle structure of the phase space makes it easy
to determine the Hilbert space of the quantum theory: it will
be given by the square integrable functions on the configuration
space. Hence in this case we shall have

\begin{equation}
{\cal H} = L^2(\R,\d p)\oplus L^2(\R,\d p).\label{111}\end{equation}

For $\Lambda<0$ neither (\ref{104}) is positive definite nor
(\ref{105}) can be obtained from (\ref{102},\ref{103}) as a
limiting case. So that the phase space is given by

\begin{equation}
T^{*}(\R^+)\>\cup \>T^{*}(\R^+)\; ,\label{106}\end{equation}
with the symplectic form
\begin{equation}\omega=2\delta\ln\lambda\delta p\; .
\label{107}
\end{equation}
The Hilbert space should be now of the form

\begin{equation}
{\cal H} =
L^2\left(\R^+,{{\d p}\over p} \right)\oplus
L^2\left(\R^+,{{\d p}\over p} \right) \; .
\label{110}\end{equation}

Although it is difficult to figure out how the Hilbert spaces
(\ref{110}) can be actually realized, we shall see in the next
sections that this prediction for the Hilbert space is consistent
with other quantization approaches.

\section{ADM formulation.}
\label{c}

In section 2 we saw explicitly that the classical solutions of
the theory are spatially homogeneous. As has been shown in
\cite{[11]} for a wide class of 2d dilaton gravity models, this
is so because the theory (\ref{action1}) has a Killing vector
the flow of which
determines a natural coordinate system on the cylinder where
the metric and the scalar fields takes an homogeneous form. The
existence of the Killing vector requires the metric equations of
motion be satisfied. At this point it is important to remark
that one can indeed reduce the theory to a finite number of
degrees of freedom by imposing the supermomentum constraint only.

To this end let us now present the basic ingredients of the ADM formulation
of the induced 2d-gravity (see also \cite{[10]}). First, we
introduce the standard parametrization of the two-dimensional
metric
\be
g_{\mu\nu}=\left(\ba{cc}-N^2 +
N_1N^1&N_1\\N_1&a^2\ea\right)\>\; ,
\label{30.1}
\ee
where $N$ and $N^1$ are the lapse and shift functions
respectively. To derive the canonical form of the action we can
use the two-dimensional identity

\be \sqrt{-g}R = -2\partial_t (aK) +
2\partial_x(a(KN^1-a^{-2}N^1))\; ,\ee
where $K$ is the extrinsic curvature scalar
\be K = \frac{1}{a^2N}(N_{1|1} -a\dot a)\; .\label{30.3}\ee
Removing total time derivatives
we arrive at

\be S = \int \d^2x(\pi_a\dot a + \pifi\dot\Phi-N{\cal C}-N^1{\cal
C}_1)\>,\label{30.4}\ee
where the canonical momenta are
\be \pia =\frac4N(\Phi'N^1-\dot \Phi)\; ,\label{30.5}\ee
\be \pifi = \frac{2a}N(\Phi'N^1-\dot \Phi)+\frac4N\left((aN^1)'-
\dot a\right)\>,\label{30.6}\ee
and the supermomentum and hamiltonian constraints are given by

\be \C_1 = \Phi'\pifi-\pia'a \; ,\label{30.7}\ee
\be \C = \frac1{16}a\pia^2-\frac14\pia\pifi-a\Lambda -
\frac1a\Phi'^2+4(a\Phi')' \; .\label{30.8}\ee

Making use of the spatial diffeomorphism invariance of the theory
we can fix the space coordinate and assume that

\be a = a(t) \; .\label{30.9}\ee
In addition to this, and due to the time reparametrization
invariance, we can also make a choice of time. All the above
considerations suggest the following class of spatially
homogeneous definitions of the internal time variable
\be {\cal T}(\Phi, a) = \chi(t) \; ,\label{30.10}\ee
where $\chi$ is a generic function. This implies
that
\be \Phi = \Phi(t) \; .\label{30.11}\ee
Now, if we impose the supermomentum constraint we easily obtain
\be \pia = \pia(t)\; ,\label{30.12}\ee
and also $N=N(t)$.

The momentum $\pifi$ is still a function of
both $t$ and $x$. However we can integrate the action in
(\ref{30.4}) with respect to the compact coordinate and the
resulting expression is
\be S = \int\d t\left(\pia\dot a + \dot \Phi\int\d x\pifi -
N\C\right)\; ,\label{30.13}\ee
where now
\be \C = \frac1{16}a\pia^2-\frac14\pia\int\d x \pifi - a\Lambda \; .
\label{30.14}\ee
{}From now on $\pifi$ stands for the momentum conjugated to
$\Phi(t)$, i.e., $\pifi(t)=\int\d x\pifi(t,x)$. Although
(\ref{30.13}) corresponds to a minisuperspace approach to the
theory, it must be regarded instead as a reduced form of the
theory in an appropriate gauge choice and not as a mere
approximation to the theory.

For the sake of completeness we write down the equations of
motion and the symplectic form obtained from (\ref{30.13})
\be\ba{ll}
\dot\pifi = 0\>,
&\dot\pia=-N\left[-\frac1{16}{\pia^2}+\frac\Lambda2\right] \; ,\\
\dot\Phi = N\pia\>,&\dot a = N\left[\frac18a\pia-
\frac14\pifi\right] \; ,\ea\label{30.15}\ee
\be \omega=\delta\pifi\delta\Phi + \delta\pia\delta a\>.
\label{omega2}\ee

% In the next two sections we shall quantize the reduced theory
% (\ref{30.13}) from both the genuine hamiltonian formalism and
% the Wheeler-DeWitt equation.

\section{Reduced phase-space quantization in the conformal
choice of time.}
In this section we shall develop a genuine hamiltonian
quantization of the reduced theory (\ref{30.13}).
In this approach the choice of time
is done before quantize and the constraint $\C = 0$ is solved
classically (see for instance the review \cite{[12]}). In this
context, the choice of time is nothing other but a gauge fixing
condition. This gauge fixing is required to be
complete in the sense that no further gauge freedom must be left, but
also we must not lose information, i.e., actual solutions to the
equation of motion.

Let us choose the conformal gauge

\be N = a\>,\label{40.1}\ee
that implies, according to the equations of motion
(\ref{30.15}) (see also (\ref{101})),
 the following implicit
definition of the time variable:
\be a^2 = 4\frac{\pifi^2}{|\Lambda|}
\frac{\e^{\pifi t}}{\left(1 -\sgn\Lambda\e^{\pifi
t}\right)^2}\>.\label{40.2}\ee
Solving now the constraint ${\cal C} =0$ for $a\pia$ we find
the solutions
\be
a\pi_a = 4\pi_\Phi\frac{1}{1+\sgn\Lambda\e^{\pifi t}}\>,
\label{40.3a}\ee
and
\be a\pi_a = 4\pi_\Phi\frac{1}{1+\sgn\Lambda\e^{-\pifi
t}}\>,\label{40.3b}\ee
which remind us the classical twofold solution for the field $\Phi$.

Once the choice of time has been done, the effective Hamiltonian
associated with it, i.e., the function that gives the proper {\it
classical} time evolution for the remaining fields, is
(minus) the conjugate momentum of time.  Substituting
(\ref{40.3a},\ref{40.3b}) into (\ref{omega2}) we find

\be \omega = \delta \pifi\delta \Phi
- 2\pi_\Phi\frac{1}{1+\sgn\Lambda\e^{\pm\pifi t}}
\delta \pifi\delta t \>.\label{omega3}\ee

Since this two-form must project down to the (reduced) symplectic
form of the model, the hamiltonian flow of the vector field in
the kernel of (\ref{omega3}) should provide the remaining
trajectories of motion (see for instance \cite{[13]}). Therefore,
the effective Hamiltonian should fit the expression

\be \omega = \delta\pifi\delta\Phi -\delta H\delta
t\>.\label{PCform}\ee
So, we obtain

\begin{eqnarray}
H &=&
\int^{\pifi}\d\pifi2\pi_\Phi\frac{1}{1+\sgn\Lambda\e^{\pm\pifi t}}
\label{Hamiltonians}\\
&=& \pifi^2 -(\pm)2\frac{\pifi}{t}\ln(1+\sgn\Lambda\e^{\pm \pifi t})
-2\frac{1}{t^2}\hbox{Polylog}(2,-\sgn\Lambda\e^{\pm\pifi
t})\>.\nonumber
\end{eqnarray}
The Hamiltonians in (\ref{Hamiltonians}) can be converted
into each other by means of the change $\pifi\leftrightarrow -\pifi$
or $t\leftrightarrow-t$. Thus, in this system, reversing the arrow
of time is equivalent to changing the sign of the momentum $\pifi$.

The quantum system will be described by the wave functions
$\Psi(\pi_\Phi,t)$ that obey a time-dependent Schr\"odinger
equation:

\be \hbox{i}\hbar\frac{\partial}{\partial t}\,\Psi(\pifi,t) =
H(\pifi,t)\Psi(\pifi,t)\>.\label{162.b}
\ee
Since the Hamiltonian functions at different times commute, the
Schr\"o-\break
dinger equation (\ref{162.b}) can be solved immediately to
give

\be \Psi(\pifi,t) = \Psi(\pifi)\e^{-\frac{i}{\hbar}\int^t\d
z\,H(\pifi,z)}\>.\label{163}\ee

The scalar  product of two wave function $\Psi(\pifi,t)$ and
$\varphi(\pifi,t)$ will be taken as the natural one:
\be
 <\Psi|\varphi>
=\int\d \pifi \Psi^*(\pifi,t)\varphi(\pifi,t)
= \int\d \pifi \Psi^*(\pifi)\varphi(\pifi)\>.\label{164}
\ee
The Hilbert space for $\Lambda>0$ is hence given by
\begin{eqnarray} {\cal H} &=& {\cal H}^{(+)}\oplus{\cal H}^{(-
)}\nonumber\\
&=&L^2(\R,\d \pifi)\oplus L^2(\R,\d \pifi)\>.\label{165}\end{eqnarray}
The two sectors correspond to the double sign of the effective
Hamiltonian (\ref{Hamiltonians}) and represent whether the
two-dimensional universe is expanding or contracting. We recover
thus the result of section 2. Note that the monodromy parameter
$a=\e^{2p}$ in section 2 must be identified with the constant
of motion $\e^{\pifi}$.

However, for $\Lambda<0$ we must prevent the wave functions to
take any non null value in $\pifi = 0$ since at this point the
gauge fixing condition (\ref{40.2}) is not well defined. So, we
must impose on the wave functions the restriction of vanishing at
$\pifi =0$,

\be \Psi(\pifi =0, t) = 0\>,\label{166}\ee
restriction that is preserved by the time evolution.
Therefore, for $\Lambda<0$, the Hilbert space will be given by

\begin{eqnarray} {\cal H}
&=& {\cal H}^{(+)}\oplus{\cal H}^{(-)}\nonumber\\
&=& L^2(\R^+,\d \pifi)\oplus L^2(\R^+,\d \pifi)\>.\end{eqnarray}
which can be identified with (\ref{110}).

\section{Quantization via the Wheeler-DeWitt equation.}

In this section we shall quantize the reduced theory
(\ref{30.13}) without any identification of time prior to
quantization. This essentially means to impose the operator
version of the classical hamiltonian constraint, i.e., the
Wheeler-DeWitt equation. To propose the Wheeler-DeWitt operator
$\C$ for ({\ref{30.14}) we face at once the problem of the
operator ordering ambiguities and the inequality $a>0$ of the
scale variable. The second difficulty can be solved by using the
affine algebra $[\hat a, \> \hat p_a]=\hbox{i}\hbar\hat a\>\> (\hat p_a =
-\hbox{i}\hbar a\frac\partial{\partial a})$, instead of the Heisenberg-
Weyl algebra, as the basic one to define the quantization
\cite{[14]}. The reason is that the operator $\hat \pia =
-\hbox{i}\hbar\frac\partial{\partial a}$ fails to be self-adjoint on
$L^2(\R^+,\d a)$, whereas the affine operator $\hat p_a =
-\hbox{i}\hbar a\frac\partial{\partial a}$ is self-adjoint
in $L^2(\R^+,\frac{\d a}a)$.

Imposing that the Wheeler-DeWitt operator be self-adjoint  with
respect to the measure $\frac{\d a}a\d \Phi$ we can write the
following expression for $\widehat \C$:
\bea \label{50.1}
\widehat{{\cal C}}&=&\Bigl[
 {1\over16} \widehat{a}^{\alpha+i\beta} \widehat{p}_a \widehat{a}^{-1-2\alpha}
\widehat{p}_a
\widehat{a}^{\alpha-i\beta} +  \nonumber \\
& & \hphantom{\Bigl[}
 -{1\over8} \left( \widehat{a}^{\gamma+i\sigma} \widehat{p}_a
\widehat{a}^{-\gamma-i\sigma-1} + \widehat{a}^{\gamma+i\sigma-1} \widehat{p}_a
\widehat{a}^{-\gamma-i\sigma} \right) \widehat{\pi}_\Phi +
 \Lambda \widehat{a} \Bigr]   \, ,
\eea
where $\alpha$, $\beta$, $\gamma$ and $\sigma$ are
arbitrary factor-ordering parameters.

We can separate variables in the Wheeler-DeWitt equation by
expanding the wave function $\Psi$ in $\widehat{\pi}_\Phi$
eigenstates

\be
\Psi = \int dq e^{{iq\Phi\over\hbar}}
\Psi_q(a) \, .
\label{50.2}
\ee
Inserting (\ref{50.2}) into the equation $\widehat\C\Psi=0$, where
$\widehat\C$ is given by (\ref{50.1}), we obtain that the
functions $\Psi_q(a)$ obey the equation

\be \label{50.3}
\left(
{d^2\over da^2} +
{1\over a}(1 - 2\zeta){d\over da} +
{1\over a^2}(\zeta^2-\nu^2) +
4 {\Lambda\over\hbar^2}
\right) \Psi_q(a) = 0 \, ,
\ee
where

\bea
\zeta &=& {1\over2} \left( 1 + 4{iq\over\hbar} + 2i\gamma \right)
        \, , \label{tresxiia} \\
\nu^2 &=& {1\over4} \left( 1- 16{q^2\over\hbar^2} - 8\gamma^2 +
4\alpha(\alpha+1) + 16 {q\over\hbar}(\sigma-\gamma) \right)
        \, . \label{tresxiib}
\eea

        The solutions of the above equation are
\be
\Psi_q(a)
= a^\zeta {\cal Z}_\nu
\left(\frac{2|\Lambda|^{\frac12}}\hbar a\right) \, ,\label{50.6}
\ee
where
${\cal Z}_\nu$ are ordinary (modified) Bessel functions for
$\Lambda>0$ ($\Lambda<0$) with order $\nu$.

        In constructing the Wheeler-DeWitt operator
we required hermiticity
with respect to the standard inner product
\be \label{50.7}
<\Psi_1|\Psi_2> = \int {da\over a} d\Phi \Psi_1^* \Psi_2 \, .
\ee
In canonical quantum gravity it is therefore natural to propose
(\ref{50.7}) as the scalar product for the solutions of the
Wheeler-DeWitt equation. This proposal for the scalar product is
problematic in the sense that we are integrating over one of the
configuration variables that could have been defined as the
``internal" time variable \cite{[12]}. However we shall insist on
using it but having in mind that (\ref{50.7}) could be divergent
and require, therefore, some sort of regularization.

Let us analyse now the situation for the case of negative
cosmological constant. We can expand the general solution to the
Wheeler-DeWitt equation in terms of the modified Bessel and
Hankel functions ${\cal I}_\nu$ and ${\cal K}_\nu$. However, due to
the exponential behaviour of the functions ${\cal I}$ for large
$x$ $(x\equiv 2\frac{|\Lambda|^{\frac12}}\hbar a)$, they do not
lead to normalizable wavefunctions and, therefore, should then be
excluded from the physical Hilbert space. The physical wave
functions should be of the form

\be
\Psi = \int dq  e^{{i q \Phi \over \hbar}} a^{\zeta}
        C(q) {\cal K}_\nu (x) \, ,
\label{50.8}
\ee

To determine the Hilbert space we should find out the range of
variation of the order $\nu$. Due to the small $x$ behaviour or the
modified Hankel functions, the wave functions will be normalizable
when
\be
\nu^2 < {1\over4} \, .
\label{50.9}
\ee
To obtain the maximum range of variation for $\nu^2$ as $q$
varies over the real line we should choose the factor ordering
parameters in such a way that (\ref{tresxiib}) turns out to be of
the form

\be
\nu^2 = {1\over4} ( 1 - 16 {(q-q_0)^2\over\hbar^2} ) \, .
\label{50.10}
\ee
The constant shift $q_0$ of $q$ in (\ref{50.10}) can be chosen
according to the classical theory. On the covariant phase space
the constant of motion $\pifi$ is proportional to the monodromy
parameter $\ln a$. Owing to the absence of classical solutions
for $a=1$ the constant $q_0$ should vanish to exclude the quantum
solution $\widehat \pifi = 0$. Therefore we are finally led to the
expression
\be
\nu^2 = {1\over4} ( 1 - 16 {q^2\over\hbar^2} ) \, ,
\label{50.11}
\ee
which corresponds to $\alpha=\beta=\gamma=\sigma=0$ in (\ref{tresxiib}).

Now we want to determine the Hilbert space when the cosmological
constant is positive. According to (\ref{50.6}) the general
solution to the Wheeler-DeWitt equation can be expanded as
($\hbox{Re}\nu\geq0\>,\>\hbox{Im}\nu\geq0)$)

\be
\Psi = \int dq a^{{1\over2}+2i{q\over\hbar}}
        e^{{iq\Phi\over\hbar}}
        \left( A(q) {\cal J}_\nu (x) + B(q)
{\cal N}_\nu (x) \right) \, ,\label{50.12}
\ee
where $A(q)$ and $B(q)$ are arbitrary complex functions and $\nu$ is
given by (\ref{50.11}). The norm of the wave function (\ref{50.12}) with
respect to (\ref{50.7}) is given by ($k=2\frac{|\Lambda|^{\frac12}}\hbar)$
\bea \label{50.13}
<\Psi|\Psi> &=& {\hbar \over k}
        \int_{-\infty}^{+\infty} dq \int_0^{+\infty} dx
  \Bigl(
        |A(q)|^2 |{\cal J}_\nu (x)|^2 +
        |B(q)|^2 |{\cal N}_\nu (x)|^2 +
 \\
 &&
\hphantom{ \pi\hbar \int_{-\infty}^{+\infty} dq }
A^*(q) B(q) {\cal J}_\nu^*(x) {\cal N}_\nu(x) +
        A(q) B^*(q) {\cal J}_\nu(x){\cal N}_\nu^*(x) \Bigr)
\nonumber\, .
\eea

Due to the asymptotic behaviour of the Bessel functions for large $x$
the above integral are divergent. We can define a regularized scalar product
by substituting the integration measure $\d x$ in (\ref{50.13}) by
$\d x/x^\epsilon$ ($\epsilon{\mathop{>}\limits_{\sim}}0)$.
In the limit $\epsilon\rightarrow0$ the new inner product turns
out to be
\bea
<\Psi|\Psi> &&{\mathop{\sim}\limits_{\scriptscriptstyle\epsilon\to0}}
        {\hbar \over 2\pi k} {\Gamma(\epsilon) \over 2^\epsilon}
        \int_{-\infty}^{+\infty} dq \Bigl\{
        \bigl[ \cos(\pi\nu) (|A|^2+|B|^2)\nonumber\\
         &&\qquad + \sin(\pi\nu) (B^* A - A^* B)
        \bigr] \Theta(-\nu^2)\label{50.14}\\
         && \qquad +
       \bigl[ |A|^2 + |B|^2 \bigr] \Theta(\nu^2) \Bigr\} \nonumber,
\eea
where $\Theta$ is the step function. We can eliminate the overall
divergent factor $\Gamma(\epsilon)/2^\epsilon$ to define the
physical scalar product. The elementary normalizable solutions
with respect to the regularized scalar product can be classified
immediately. They are
$\J_\nu$ for $\nu\in[0,\frac12]$ or $\Re\>\nu=0$,
and
$\N_\nu$ for $\nu\in[0,\frac12[$ or $\Re\>\nu=0$.
Note that the unique normalizable solution for $\nu=\frac12$ is
$\J_\nu$.

Next we would like to relate the quantization obtained via the
Wheeler-DeWitt equation with the approach developed in previous
sections. The main point is to see how the Hilbert space
$L^2(\R^+)\oplus L^2(\R^+)$ (or $L^2(\R)\oplus L^2(\R)$,
depending on the sign of $\Lambda$), obtained from the covariant
and reduced phase-space quantizations, can be realized in terms of the
normalizable solutions of the Wheeler-DeWitt equations. Let us
first consider the case of negative cosmological constant. Any
normalizable solutions $\Psi$ of the form (\ref{50.8}) can be
decomposed as
$\Psi = \Psi^{(+)} + \Psi^{(-)}$, where (we have redefined the
function $C(q)$)
\bea
\Psi^{(+)} &=& \left( {k\over\pi\hbar}
                 \Gamma({1\over2}+\nu) \Gamma({1\over2}-\nu)
                \right)^{1\over2}
                \int_0^\infty dq a^{{1\over2}+2i{q\over\hbar}}
                e^{iq{\Phi\over\hbar}} C^{(+)}(q)
                {\cal K}_{|\nu|} (x)
                \label{50.15} \, , \\
\Psi^{(-)} &=& \left( {k\over\pi\hbar}
                 \Gamma({1\over2}+\nu) \Gamma({1\over2}-\nu)
                \right)^{1\over2}
                \int^0_{-\infty} dq a^{{1\over2}+2i{q\over\hbar}}
                e^{iq{\Phi\over\hbar}} C^{(-)}(q)
                {\cal K}_{|\nu|} (x)
                \label{50.16} \, .
\eea

The scalar product takes the form
\be
< \Psi | \Psi > =
\int_0^\infty dq |C^{(+)}(q)|^2 +
\int^0_{-\infty} dq |C^{(-)}(q)|^2
\label{50.17} \, ,
\ee
and this shows the coincidence with the Hilbert space
derived in sections 2 and 4.

When the cosmological constant is positive, the Hilbert space of
normalizable solutions of the Wheeler-DeWitt equation can also be
decomposed into two orthogonal subspaces. Any normalizable
solution $\Psi$ of the form (\ref{50.12}) can be split as
$\Psi = \Psi^{(+)} + \Psi^{(-)}$, where

\bea
\Psi^{(+)} &=& \left({k\over2\hbar}\right)^{1\over2}
                \int_{-\infty}^{+\infty} dq
                a^{{1\over2}+2i{q\over\hbar}} e^{iq{\Phi\over\hbar}}
                A^{(+)}(q)
                \Bigl[ \Theta(q) {\cal J}_\nu (x)  \nonumber \\
                &&\hphantom{\left({k\pi\over\hbar}\right)^{1\over2}
                                \int_{-\infty}^{+\infty} dq}
                       + \Theta(-q) \bigl(
                                (\cos(\pi Im(\nu))^{1\over2}
                                {\cal N}_\nu(x)  \nonumber \\
                && \hphantom{\left({k\pi\over\hbar}\right)^{1\over2}
                                \int_{-\infty}^{+\infty} dq}
                              +  \sin(\pi Im(\nu))
                                \left(cos(\pi Im(\nu))\right)^{-{1\over2}}
                                {\cal J}_\nu (x)
                        \bigr)
                \Bigr] \, , \label{50.18} \\
\Psi^{(-)} &=& \left({k\over2\hbar}\right)^{1\over2}
                \int_{-\infty}^{+\infty} dq
                a^{{1\over2}+2i{q\over\hbar}} e^{iq{\Phi\over\hbar}}
                A^{(-)}(q)
                \Bigl[ \Theta(-q) {\cal J}_\nu (x)  \nonumber \\
                && \hphantom{\left({k\pi\over\hbar}\right)^{1\over2}
                                \int_{-\infty}^{+\infty} dq}
                       + \Theta(q) \bigl(
                                (\cos(\pi Im(\nu))^{1\over2}
                                {\cal N}_\nu(x)  \nonumber \\
                && \hphantom{\left({k\pi\over\hbar}\right)^{1\over2}
                                \int_{-\infty}^{+\infty} dq}
                              +  \sin(\pi Im(\nu))
                                \left(cos(\pi Im(\nu))\right)^{-{1\over2}}
                                {\cal J}_\nu (x)
                                \bigr)
                \Bigr] \, . \label{50.19}
\eea
The resulting expression for the scalar product turns out to be

\be
< \Psi | \Psi > =
\int_{-\infty}^{+\infty} dq \left( |A^{(+)}(q)|^2 + |A^{(-)}(q)|^2 \right)
\, . \label{50.20}
\ee
and shows the equivalence between the Hilbert space derived from
the Wheeler-DeWitt equation and the one obtained from the covariant
and reduced phase-space approach.

\section{Final comments}

In this paper we have constructed the quantum theory of the induced 2d-gravity
in three different ways: i) The covariant phase space quantization; ii) The
reduced ADM phase-space quantization and iii) The reduced Wheeler-DeWitt
equation. We have explicitly shown the coincidence of the Hilbert space of
these approaches. The first approach is based on the space of classical
solutions and it permits to determine the ``size" of the Hilbert space.
The other approaches lead to two different realizations of the Hilbert
space. The comparison between the different approaches allows to
understand the role played by the classical solutions in the quantum theory.
The absence of normalizable wavefunctions for $\nu^2=\frac14$, when
$\Lambda<0$, can be understood as the absence of classical parabolic solutions.
Furthermore, the existence of an unique wavefunction for $\nu^2=\frac14$, when
 $\Lambda>0$, find its classical counterpart in the existence of an unique
classical parabolic solution (up to and additive constant for the scalar
field).

Finally we want to remark that one could obtain an inequivalent quantization
if both the hamiltonian and the supermomentum constraints were imposed at the
quantum level. Solving the supermomentum constraint classically prevent the
emergence of the ``Schwinger term" in the algebra of surface deformations
generated by ${\cal C}_1$ and ${\cal C}$ (the central extension involves both
hamiltonian and supermomentum constraints \cite{[2]}).

\section*{Acknowledgements}

        M. Navarro acknowledges to the MEC for a Postdoctoral fellowship.
C. F. Talavera is grateful to the {\it Generalitat Valenciana} for a FPI grant.


\begin{thebibliography}{99}

\bibitem{[1]}
        R. Jackiw, in {\it Quantum Theory of Gravity}, ed. S. Christensen
        (Adam Hilger, Bristol, 1984) p. 403.
\bibitem{[2]}
        C. Teitelboim, in {\it Quantum Theory of Gravity}, ed. S. Christensen
        (Adam Hilger, Bristol, 1984) p. 327.
\bibitem{[3]}
        A. M. Polyakov, {\it Phys. Lett.} {\bf B163} (1981) 207;
        {\it Mod. Phys. Lett.} {\bf A} 2 (1987) 899.

\bibitem{[4]}
        J. Navarro-Salas, M. Navarro and V. Aldaya, {\it Nucl. Phys.}
        {\bf B} 403 (1993) 291.

\bibitem{[5]}
        C. Crnkovic and E. Witten, in {\it Three Hundred Years of
        Gravitation}, eds. S. W. Hawking and W. Israel
        (Cambridge, Cambridge, 1987) p. 676.
\bibitem{[6]}E. Witten, {\it Nucl. Phys.} {\bf B311} (1988)46.

\bibitem{[7]} K. Gawedzki, {\it Comm. Math. Phys.} {\bf
139}(1991)201.

\bibitem{[8]}
        J. Navarro-Salas, M. Navarro and V. Aldaya, {\it Phys.
Lett.} {\bf B}287 (1992)109-118.

\bibitem{[9]}
        A. Mikovic, Imperial College preprint {\bf TP}/92-93/15.

\bibitem{[11]}T. Banks and M. O'Loughlin, {\it Nucl. Phys.}
{\bf B362} (1991)649-664.

\bibitem{[10]}
        C. G. Torre, {\it Phys. Rev.} {\bf D} 40 (1989) 2588.


\bibitem{[12]}C.J. Isham, "Conceptual and Geometrical Problems
in Quantum Gravity", Imperial/{\bf TP}/90-91/14.

\bibitem{[13]} R. Abraham and J. Marsden, ``Foundations of
Mechanics", Benjamin, New York, (1978).\\
V. Aldaya and J. A. de Azcarraga,
{\it La Rivista del Novo Cimento} 3, 3, 10 (1980).

\bibitem{[14]}
        J. R. Klauder and E. Aslaksen, {\it Phys. Rev.} {\bf D} 2
        (1970) 393.\\
        C. J. Isham, in {\it Relativity, Groups and Topology
        II}, Proc. 1983 { Les Houches Summer School}, ed. B. DeWitt and
        R. Stora (North-Holland, Amsterdam, 1984).



\end{thebibliography}
\end{document}